\documentclass[11pt]{article}
\usepackage{epsfig}
\newcommand{\nb}{\nonumber}
\newcommand{\vpar}{\vskip2mm \par}

\begin{document}

\title{Probing R-parity violating interactions via $p\bar{p} \to e\mu+X$ channel on Tevatron
       \footnote{Supported by National Natural Science Foundation of China and special fund
       sponsored by China Academy of Science.}}

\vspace{3mm}

\author{{ SUN Yan-Bin$^{b}$, JIANG Yi$^{b}$, HUANG Jin-Rui$^b$,
          HAN Liang$^{b}$, ZHANG Ren-You$^{b}$, MA Wen-Gan$^{a,b}$}\\
      {\small $^{a}$ CCAST (World Laboratory), P.O.Box 8730, Beijing 100080, P.R.China} \\
      {\small $^{b}$ Department of Modern Physics, University of Science and Technology}\\
      {\small of China (USTC), Hefei, Anhui 230026, P.R.China}\\
% {\small Email: sunyb@mail.ustc.edu.cn, jiangyi@ustc.edu.cn, huangjr@mail.ustc.edu.cn,}\\
% {\small \hskip 13mm hanl@ustc.edu.cn, zhangry@ustc.edu.cn, mawg@ustc.edu.cn}
}

\date{}
\maketitle

\begin{abstract}
We investigated the lepton flavor violation processes $p\bar{p}
\to e\mu+X$ induced by R-parity violating interactions at the
Tevatron hadron collider. The theoretical calculation and Monte
Carlo simulation demonstrate that with a set of suitable cuts on
experimental observables, one might be capable to reduce the
standard model physical background to a controllable level so that
the signals of R-parity violating interactions could be detected
distinctively. Furthermore, clear sneutrino information could be
abstracted from the purified event sample where other SUSY scalar
quark 'pollution' is heavy suppressed. We conclude that with a
reasonable assumption of $10fb^{-1}$ integrated luminosity, the
Experiments at the Tevatron machine would have potential to
discover sneutrino in the region of $m_{\tilde{\nu}}\le400GeV$ via
lepton flavor violation $e\mu$ production channels, or extend the
mass scale constraint up to $m_{\tilde{\nu}}\ge 550~GeV$ at 95\%
CL.
\end{abstract}

\vskip 20mm

%11.30.Fs   Global symmetries (e.g., baryon number, lepton number)
%11.30.Pb   Supersymmetry (see also 12.60.Jv Supersymmetric models)
%12.60.Jv   Supersymmetric models (see also 04.65.+e Supergravity)
%14.80.Ly   Supersymmetric partners of known particles

%\keywords{Lepton Flavor Violation, the R-violating Minimal Supersymmetric Standard Model, Hadron Collider}

\vskip 5mm

\section{Introduction}

The observed neutrino oscillation\cite{neutrinoexp} implies strong
lepton flavor violation(LFV). In the Standard Model(SM), lepton
number is exactly preserved in contradiction with the neutrino
oscillation. Some shortcomings of SM, for instance, the emergence
of quadratic divergences in the Higgs sector, imply that the SM
need to be extended. The Minimal Supersymmetric Standard Model
(MSSM)\cite{MSSM} is one of the most promising candidates in all
extensions of the SM. The most general superpotential in
supersymmetry theory contains bilinear and trilinear terms which
do not conserve either the baryon number (B) or the lepton number
(L). The simultaneous presence of both lepton number violating and
baryon number violating could lead to very rapid proton decay. In
order to turn off proton rapid decay, supersymmetric models
introduce a discrete R-parity symmetry\cite{R} implying a
conserved quantum number $R=(-1)^{3B+L+S}$, where B, L and S are
the baryon number, the lepton number and the spin of particles
respectively. However, such a stringent symmetry appears short of
theoretical basis. Especially we know that a stable proton can
survive by imposing only one of L- and B- conservation. Moreover,
non-zero R-violating couplings might provide small neutrino
masses, which would explain the phenomena of neutrino oscillation
experiments. Thus, there is strong theoretical and
phenomenological motivation to introduce partial R-parity
violations into the most general representations of
superpotential, which can be written as \cite{potential1}
\begin{equation}
\label{sup} {\cal W}_{\rlap/R_{p}} = \frac{1}{2}\epsilon_{ab}
\lambda_{ijk}\hat{L}_{i}^a \hat{L}_{j}^b \hat{E}_{k} +
\epsilon_{ab}\lambda^{'}_{ijk} \hat{L}_{i}^a \hat{Q}_{j}^b
\hat{D}_{k} +
\frac{1}{2}\epsilon_{\alpha\beta\gamma}\lambda^{''}_{ijk}
   \hat{U}_{i}^{\alpha} \hat{D}_{j}^{\beta} \hat{D}_{k}^{\gamma} +
\epsilon_{ab}\delta_{i} \hat{L}_{i}^a \hat{H}_{2}^b
\end{equation}
where $i,j,k$ = 1,2,3 are generation indices; $a,b$ = 1,2 are
SU(2) isospin indices and $\alpha,\beta,\gamma$ are SU(3) color
indices. $\lambda,\lambda^{\prime},\lambda^{\prime\prime}$ are
dimensionless R-violating Yukawa couplings behaving as
$\lambda_{ijk}=-\lambda_{jik}$,
$\lambda^{\prime\prime}_{ijk}=-\lambda^{\prime\prime}_{jik}$. In
above superpotential, the last bilinear terms mix the lepton
superfield and the Higgs one, which might generate masses of
neutrinos and consequently introduce compatible description of
neutrino oscillation in a natural way\cite{Rbilinear}. All the
other trilinear terms only violate either L- or B-symmetry
respectively, and the terms that may produce both L- and
B-violation simultaneously are forbidden in superpotential so that
a stable proton is ensured.

\par
Experimental detection on the signals of R-parity violating
interactions would be the most outstanding evidence of new physics
beyond the SM. If nature is supersymmetric and neutrino
oscillations are really induced by R-violating couplings included in a
general superpotential, then the prediction of single sneutrino
production modes at colliders is straightforward.  Therefore,
probing single sneutrino production on a high energy collider is
specially attractive in experimental searches and relevant
phenomenological studies about R-violating interactions.  Such
exploration ought to be promising if carried out on fine high
energy lepton colliders. There were some experimental bounds on
LFV di-lepton cross-section measurement from LEP
experiments\cite{LEPOPAL}, which could attribute to single
sneutrino productions and would set mass constraint on these
neutrino super-partners. We have discussed the possibility of
probing off-shell sneutrino effect induced by $\hat L \hat L \hat
E$ terms on a 500GeV Linear Collider \cite{eeemJHEP}, and
concluded that with current constraints on $\lambda$ parameters
one can extend sneutrino search up to $m_{\tilde \nu} \ge 1$TeV
region which is far beyond the constraints from LEP. Nevertheless,
despite the high performance of Linear Colliders and clean
environment in their detectors, none of TESLA, JLC or even ILC
projects would be in commission in this decade. On the other hand,
the Tevatron hadron collider is currently in its upgrade Run-II
physics and the LHC will be put into running in three years, and
consequently more careful studies on direct SUSY searches
including the potential detection of single sneutrino production
on hadron colliders should be addressed in detail both
experimentally and phenomenologically.

\par
The first two terms $\hat{L}\hat{L}\hat{E}$ and
$\hat{L}\hat{Q}\hat{D}$ in the superpotential Eq.(\ref{sup}) may
lead to R-violating single sneutrino production and sequential LFV
decay on hadron colliders. Many papers studied these single
sneutrino production modes both on- and off-mass-shell of
sneutrino resonance via $\hat{L}\hat{Q}\hat{D}$ interactions and
decay in R-parity conservation mode as $\tilde{\nu}\rightarrow
l\tilde{\chi}^{\pm}_{i}$ \cite{sneu}. Charginos and subsequential
neutrolinos could decay in R-conserving way as well, e.g.
$\tilde{\chi}^{\pm}_{i}\rightarrow \tilde{\chi}^{0}_{j} W^{*}
\rightarrow \tilde{\chi}^{0}_{j} l \nu$ and
$\tilde{\chi}^{0}_{j}\rightarrow l \tilde{l}$; then sleptons from
neutrolino would decay in R-violating modes as $\tilde{l}
\rightarrow l\nu$ or $\tilde{l} \rightarrow u d$. With these
R-violating interaction vertexes involved in sneutrino and
chargino/neutrolino decay chains, single sneutrino will
subsequently decay to all the SM stable particles at last. Namely,
these single sneutrino production and subsequential
$\chi^\pm/\chi^0$ decay processes are doubly R-violated, and
unless light neutrolinos would live long enough to escape from
detectors there would be more than three leptons in final states.
These characteristic triple-lepton signals are obviously good for
tagging new physics. But there are problems in SUSY parameter
determination from these multiple leptons channels: Firstly, the
tri-lepton production can also be induced by R-conserving
supersymmetric models, for instance,
$\tilde{\chi}^{\pm}_{i}\tilde{\chi}^{0}_{j}$ associated production
may have the subsequential decay to three leptons and two stable
$\tilde{\chi}^{0}_{1}$ as the lightest supersymmetric particle
(LSP) which performances as missing energy in detector. So, these
tri-lepton channels might not be able to distinguish
supersymmetric R-violating interaction signals from those of
R-conserving ones. Secondly, since there are too many parameters
involving in $\tilde{\nu}\rightarrow l\tilde{\chi}^{\pm}_{i}$ and
chargino and neutrolino cascade decay, it is complicated to
determine any of SUSY parameters decisively, and one has to adopt
a very compactly simplified version of parameters such as mSUGRA.
On the contrary, the LFV di-lepton processes of sneutrino rare
decay $\tilde{\nu}\rightarrow l^-_i l^+_j$ on hadron collider
would provide an ideal detection channel for R-violating
interactions and a potential means of abstracting some SUSY
parameters experimentally. As we will show in following sections,
with some reasonable assumption on parameter space and appropriate
treatment on experimental observables, one is not only able to
select LFV di-lepton signals from numerous SM physics background,
but also able to decouple the contribution of sneutrino part in
signals from that of other SUSY parts, so that an unique sneutrino
mass parameter could be abstracted explicitly.

\par
Leaving alone the LHC in construction, Tevatron Run-II upgrade is
now in progress. It is expected to deliver an proton-antiproton
collision integrated luminosity as $\cal O$($fb^{-1}$) per
experiment at the energy of 1.96 TeV , which will be
quantitatively one order higher than the luminosity acquired in
Run-I with center of mass energy 1.8 TeV. Therefore, before LHC,
the two Experiments D\O\ and CDF on the Tevatron are the only
available facilities to test R-violating interaction and search
for sneutrino. In this paper, we will discuss the possibility of
probing R-violating LFV di-lepton signals and abstracting
sneutrino information from experimental observation at the
Tevatron Run-II.

\vskip 5mm
\section{Analytical expression on the $e\mu$ signal}

We investigate two di-lepton signal processes in which final
states can be identified precisely even in hadron colliding
environment, namely  $p\bar p\to q\bar q/\bar q q\to e^-\mu^+$ and
its charge-conjugated process $p\bar p\to \bar q q/q \bar q \to
e^+\mu^-$. By integrating the first two lepton number violation
terms in  Eq.(\ref{sup}), one can obtain the lagrangian relevant
to present discussion as
\begin{eqnarray}
\label{lag} {\cal L}_{\rlap /L}&=& \frac{1}{2} \lambda_{ijk}\cdot
(\bar \nu_{Li}^c e_{Lj} \tilde e_{jL}^* + e_{Li} \bar \nu_{Lj}^c
\tilde e_{Rk}^* + \nu_{Li} e_{Lj} \bar e_{Rk} - e_{Li} \tilde
\nu_{Lj} \bar e_{Rk}) + \nb\\
&&~~\lambda_{ijk}^{'}\cdot(\bar \nu_{Li}^c
d_{Lj} \tilde d_{Rk}^* - e_{Ri}^c u_{Lj} \tilde d_{Rk}^* +
\nu_{Li} \tilde d_{Lj} \bar d_{Rk} - e_{Li} \tilde u_{Lj} \bar
d_{Rk} +\nb\\
&&~~~~~~~~~~\tilde \nu_{Li} d_{Lj} \bar d_{Rk} - \tilde e_{Li} u_{Lj}
\bar d_{Rk})~+~h.c.
\end{eqnarray}
where superscript $c$ refers to charge conjugation. In this work,
we simply take the R-violating parameters $\lambda$ and
$\lambda^{'}$ to be real to avoid further complication.

\par
In order to get optimal efficiency of signal detection, we choose
not to distinguish the absolute sign of lepton charge, so that
di-track events of $e^-\mu^+$ final states would be treated
equally as those of $e^+\mu^-$ ones. This summation treatment of
$e\mu$ events\footnote{The notation of charge-unspecified $e\mu$
denotes the summation of two final states $e^-\mu^+$ and
$e^+\mu^-$ events.} is based on the following consideration:
Firstly, it can double the statistic of signal events, which will
improve the significance and confidence on collected signal
samples. Secondly, the charge determination of tracks with high
transverse momentum($p_T$) relies on many realistic
detector-relevant factors, thus indiscrimination between
$e^-(\mu^+)$ and $e^+(\mu^-)$ will remove systematic uncertainty
on charge measurement. To reflect the summation measurement in
theoretical calculation, we phenomenologically introduce an unique
parton-level $\hat {\theta}$, to denote polar angles of both
outgoing $e^-$ and $e^+$ with respect to the incoming parton from
proton beam, and adopt a consistent momentum notation to all
signal processes as
\begin{eqnarray}
\label{proc1}
 q(p_1)+\bar q(p_2)\rightarrow e^-(p_3)+\mu^+(p_4) \\
\label{proc2}
 q(p_1)+\bar q(p_2)\rightarrow e^+(p_3)+\mu^-(p_4) \\
\label{proc3}
 \bar q(p_1)+q(p_2)\rightarrow e^-(p_3)+\mu^+(p_4) \\
\label{proc4}
 \bar q(p_1)+q(p_2)\rightarrow e^+(p_3)+\mu^-(p_4)
\end{eqnarray}
where $q=u,d$ are the first generation partons,
$p_1$ stands for the four-momenta of partons
from proton beam whose direction of vector $\vec{p}_1$ is along
$\hat z$ axis, and $p_2$ stands for those from anti-proton beam. We
always use the four-momentum $p_3$ to denote outgoing $e^-$
and $e^+$ in signal processes, so that the open angle between
$\vec p_3$ and $\vec p_1$ is precisely $\hat \theta$ defined
above. Correspondingly, $p_4$ represents the four-momenta of final
$\mu^+(\mu^-)$ particles. Neglecting electron and muon masses, we
introduce Mandelstam invariant variables to describe kinematics of
all the four signal subprocesses as
\begin{eqnarray}
 \hat s &=& (p_1 + p_2)^2=(p_3 + p_4)^2\nonumber\\
 \hat t &=& (p_1 - p_3)^2=(p_2 - p_4)^2 = -\frac{\hat s}{2}(1-\cos\hat
 {\theta})\\
 \hat u &=& (p_1 - p_4)^2=(p_2 - p_3)^2 = -\frac{\hat s}{2}(1+\cos\hat
 {\theta})
\nonumber
\label{mandelstam}
\end{eqnarray}
and Feynman diagrams of $e\mu$ signals at parton level are
depicted in Figure 1 accordingly.
\par
We define the amplitude of $d \bar d \rightarrow e^-\mu^+$
subprocess as ${\cal M}_{d\bar d}^{(-)}$, where the superscript
minus sign in parenthesis indicates the processes with negative
charged electrons in final states, and in the following we will
use the superscript $(+)$ to denote processes with positive
charged outgoing positrons. The amplitude ${\cal M}_{d\bar
d}^{(-)}$ can be represented by $\hat s$- and $\hat t$-channel
parts
\begin{equation}
{\cal M}_{d\bar d}^{(-)} = {\cal M}^{(-) \hat s}_{d\bar d} - {\cal
M}^{(-) \hat t}_{d\bar d}.
\end{equation}
with
\begin{eqnarray}
\label{amps} {\cal M}^{(-) \hat s}_{d\bar d} &=& \sum_{i=1}^3
  i~\lambda_{i12} \lambda^{'}_{i11}\cdot
  \bar u(p_3)P_R v(p_4)\cdot {\cal P}(\hat s,m_{\tilde
  \nu_i})\cdot{\bar v}(p_2)P_L u(p_1) \nb\\
&~~~& +~(P_L \leftrightarrow P_R,~\lambda_{i12}
\lambda^{'}_{i11}\rightarrow \lambda_{i21} \lambda^{'}_{i11}),
\end{eqnarray}
\begin{eqnarray}
\label{ampt} {\cal M}^{(-) \hat t}_{d\bar d} =
\sum_{i=1}^3\sum_{k=1}^2
  -i~\lambda^{'}_{1i1}\lambda^{'}_{2i1}|R^{\tilde u_i}_{1k}|^2\cdot
  \bar v(p_2) P_L v(p_4) \cdot {\cal P}(\hat t,m_{\tilde u_{i,k}}) \cdot
   \bar u(p_3)P_R u(p_1), \nb \\
\end{eqnarray}
and the notations of propagator factors ${\cal P}(x,m)$ are
defined as follows
\begin{eqnarray}
\label{propg}
{\cal P}(\hat t,m) &=& \frac{1}{\hat t - m^2}, \nonumber\\
{\cal P}(\hat u,m) &=& \frac{1}{\hat u - m^2}, \nonumber\\
{\cal P}(\hat s,m) &=& \frac{1}{\hat s - m^2+i \hat s \Gamma/m},
\end{eqnarray}
where a sparticle width $\Gamma$ is introduced into the $\hat
s$-channel propagators to suppress resonance singularity.
According to the Feynman diagrams in Figure 1, the amplitude
of $u \bar u \rightarrow e^-\mu^+$ subprocess has only
$\hat u$-channel part and can be denoted as
\begin{equation}
{\cal M}_{u\bar u}^{(-)} = {\cal M}^{(-) \hat u}_{u\bar u}, \nb
\end{equation}
with
\begin{eqnarray}
\label{ampu} {\cal M}^{(-) \hat u}_{u\bar u} =
~\sum_{i=1}^3\sum_{k=1}^2
-\frac{i}{2}\lambda^{'}_{11i}\lambda^{'}_{21i}|R^{\tilde
  d_i}_{2k}|^2\cdot \bar u(p_4)P_L u(p_1)\cdot{\cal P}(\hat
u,m_{\tilde d_{i,k}})\cdot\bar v(p_2)P_Rv(p_3). \nb \\
\end{eqnarray}
In above equation, $P_{L/R}$=$(1\mp \gamma_{5})/2$ are left- and
right-hand project operators. $i(=1,2,3)$ is generation index of
supersymmetric particles, $k(=1,2)$ denotes the two mass
eigenstates of each scalar quark flavor. $R^{\tilde u_i}$ and
$R^{\tilde d_i}$ are the $2 \times 2$ matrices used to diagonalize
various up and down-type squark mass matrices, respectively.

\par
The amplitudes of the other type of signal subprocess
$q\bar{q}\rightarrow e^+\mu^-$, represented by Eq.(\ref{proc2}), can
be written out straightforward from above expressions of
Eq.(\ref{proc1}) as
\begin{eqnarray}
{\cal M}^{(+)\hat s}_{d\bar d} &=& {\cal M}^{(-)\hat s}_{d\bar
  d}(p_3\leftrightarrow p_4) \nb \\
{\cal M}^{(+)\hat u}_{d\bar d} &=& {\cal M}^{(-)\hat t}_{d\bar
  d}(p_3\leftrightarrow p_4, \hat t \rightarrow \hat u) \\
{\cal M}^{(+)\hat t}_{u\bar u} &=& {\cal M}^{(-)\hat u}_{u\bar
  u}(p_3\leftrightarrow p_4, \hat u \rightarrow \hat t) \nb
\end{eqnarray}

\par
The amplitudes of initial first generation sea quark collision subprocesses,
represented by Eq.(\ref{proc3},\ref{proc4}), can be obtained from
those of valence incoming quark subprocesses of
Eq.(\ref{proc1},\ref{proc2}) under following replacement
\begin{eqnarray}
{\cal M}^{(\mp)\hat s}_{\bar q q} &=& {\cal M}^{(\mp)\hat
s}_{q\bar q}(p_1\leftrightarrow p_2) \nb \\
{\cal M}^{(\mp)\hat u}_{\bar q q} &=& {\cal M}^{(\mp)\hat
t}_{q\bar q}(p_1\leftrightarrow p_2,\hat t\leftrightarrow \hat u) \\
{\cal M}^{(\mp)\hat t}_{\bar q q} &=& {\cal M}^{(\mp)\hat
u}_{q\bar q}(p_1\leftrightarrow p_2,\hat u\leftrightarrow \hat t)
\nb
\end{eqnarray}
By now all the signal amplitudes at parton level are given, where
those of subprocess (\ref{proc1}) are written out explicitly,
while those of the other three ones are obtained by appropriate
initial and final state exchange respectively.

\par
In this paper, we mainly focus on probing sneutrino resonance
effect via potential LFV $e\mu$ signals at the Tevatron Run-II.
Since the $\hat t$ or $\hat u$-channel squark internal exchange
diagrams contribute into signal measurement and arose ambiguity in
exploring sneutrino information, which makes sense via the $\hat
s$-channel exchange, we name all the squark contribution as
'pollution' of the signal. Fortunately, because of the R-violating
scalar-pseudoscalar(S-P) Yukawa couplings of sfermion to Dirac
fermions as shown in Eq.(\ref{lag}), the interference terms among
$\hat s$-channel sneutrino exchange diagrams and $\hat t$ and
$\hat u$-channel squark interchange ones get vanished in all
subprocesses. We call the vanishing of the interference terms
between amplitudes of ${\cal M}^{(\mp)\hat s}$ and those of ${\cal
M}^{(\mp)\hat t+ \hat u}$, as a 'decouple' feature between the two
parts of sneutrino and squark contributions. In this way, we might
be able to abstract sneutrino information from
$p\bar{p}\rightarrow e\mu$ measurement. In order to reduce the
number of parameters in sneutrino and squark sections, we simply
take that the mass spectrum of all flavor scalar quarks is highly
degenerated, and reasonably assume the mass splitting among three
heavy sneutrinos is trivial so that an unique mass scale parameter
could be introduced in the sneutrino section. That means
\begin{equation}
m_{{\tilde u}_{i,k}} = m_{{\tilde d}_{i,k}} = m_{\tilde q}
\end{equation}
\begin{equation}
m_{\tilde\nu_i} = m_{\tilde\nu}
\end{equation}

\par
To reflect the $e\mu$ summation and the decouple feature between
sneutrino and squark contributions in the calculation of the
signal production cross-section at hadron-level, we define a set
of parton-level differential cross-section components as
\begin{eqnarray}
d\hat {\sigma}^{\hat{s}}_{qq^{'}} &=& \frac{1}{4}\frac{1}{9}
\frac{N_c}{32\pi \hat s} \cdot |{\cal M}^{\hat s}_{qq^{'}}|^2
\cdot d\cos\hat \theta,
~~~~ (qq^{'}=d\bar{d},\bar{d}d)  \\
\nb\\
d\hat {\sigma}^{\hat{t}+\hat{u}}_{qq^{'}} &=&
\frac{1}{4}\frac{1}{9} \frac{N_c}{32\pi \hat s} \cdot |{\cal
M}^{\hat{t}+\hat{u}}_{qq^{'}}|^2 \cdot d\cos\hat \theta, ~~~~
(qq^{'}=u\bar{u},\bar{u}u,d\bar{d},\bar{d}d)
\end{eqnarray}
where
\begin{eqnarray}
\label{Msdd}
|{\cal M}^{\hat s}_{d\bar d}|^2 &=& |{\cal M}^{\hat
s}_{\bar d d}|^2 = |{\cal M}^{(-)\hat s}_{d\bar{d}}|^2+|{\cal
M}^{(+)\hat s}_{d\bar d}|^2
 \nb \\
&=& 2\times \hat s^2 \cdot
 |{\cal P}(\hat s,m_{\tilde \nu})|^2 \cdot \nb\\
&&\sum_{i,i^{'}=1}^3 \lambda^{'}_{i11}\lambda^{'}_{i^{'}11}
(\lambda_{i12}\lambda_{j12}+\lambda_{i21}\lambda_{j21}) \\
\nb\\
\label{Mtudd}
|{\cal M}^{\hat t+\hat u}_{d\bar d}|^2 &=& |{\cal
M}^{\hat t +\hat u}_{\bar d d}|^2= |{\cal M}^{(-)\hat
t}_{d\bar{d}}|^2+|{\cal M}^{(+)\hat u}_{d\bar d}|^2 \nb \\
&=& [~\hat u^2\cdot {\cal P}(\hat u,m_{\tilde q})^2+\hat
       t^2\cdot {\cal P}(\hat t,m_{\tilde q})^2~]\cdot \nb \\
&&\sum_{i,i^{'}=1}^{3}\sum_{k,k^{'}=1}^{2}(~\lambda^{'}_{1i1}
   \lambda^{'}_{2i1}|R_{k1}^{\tilde u_i}|^2\cdot
 \lambda^{'}_{1i^{'}1}\lambda^{'}_{2i^{'}1}|R_{k^{'}1}^{\tilde
       u_{i^{'}}}|^2~) \\
\nb\\
\label{Mtuuu}
|{\cal M}^{\hat t+\hat u}_{u\bar u}|^2 &=& |{\cal
M}^{\hat t +\hat u}_{\bar u u}|^2 = |{\cal M}^{(-)\hat
u}_{u\bar{u}}|^2+|{\cal
M}^{(+)\hat t}_{u\bar u}|^2 \nb \\
&=& [~\hat u^2\cdot {\cal P}(\hat u,m_{\tilde q})^2+\hat
       t^2\cdot {\cal P}(\hat t,m_{\tilde q})^2~]\cdot \nb\\
   && \sum_{i,i^{'}=1}^3\sum_{k,k{'}=1}^2
       (~\lambda^{'}_{21i}\lambda^{'}_{11i}|R_{k2}^{\tilde d_i}|^2 \cdot
          \lambda^{'}_{21i^{'}}\lambda^{'}_{11i^{'}}
       |R_{k^{'}2}^{\tilde d_{i^{'}}}|^2~)
\end{eqnarray}
The above expressions manifest that with the summation treatment
on $e\mu$ signal processes, the $\hat s$-channel contribution is
doubled, while $\hat t$ and $\hat u$-channel contributions always
come together and will cancel their individual forward-backward
distribution asymmetry on polar angle $\hat \theta$ with each
other at parton level.

\vpar
Due to the decouple feature between sneutrino and squark
sections, we are safe to divide the calculation of $e\mu$
signal total cross-section at the Tevatron into two separate parts
named as sneutrino and squark cross-section contributions,
respectively as
\begin{equation}
\label{equation 1}
 \sigma[p\bar p \rightarrow e\mu] =
\sigma^{(\tilde{\nu})} + \sigma^{(\tilde{q})}
\end{equation}
And the differential cross-sections are
\begin{eqnarray}
\label{equation 2}
\frac{d\sigma^{(\tilde{\nu})}}{dx_1dx_2d\cos\hat \theta}=
\sum_{qq^{'}=d\bar{d},\bar{d}d} f_q^{p}(x_1,Q^2)f_{q^{'}}^{\bar
p}(x_2,Q^2)
    \frac{d\hat{\sigma}^{\hat s}_{qq^{'}}(\hat s =x_1 x_2
    s)}{d\cos\hat\theta} \nb\\
\end{eqnarray}
and
\begin{eqnarray}
\label{equation 3} \frac{d\sigma^{(\tilde{q})}}{dx_1dx_2d\cos\hat
\theta}= \sum_{qq^{'}=u\bar{u},\bar{u}u \atop
        ~~~~~d\bar{d},\bar{d}d}
f_q^p(x_1,Q^2)f_{q^{'}}^{\bar p}(x_2,Q^2)
    \frac{d\hat{\sigma}^{\hat t+\hat u}_{qq^{'}}(\hat s =x_1 x_2
    s)}{d\cos\hat\theta} \nb\\
\end{eqnarray}
We adopt the Pythia \cite{pythia} built-in parton distribution
function (p.d.f) of the proton and antiproton, CTEQ5L,
and implement above hadron-level calculation by convoluting
Eq.(\ref{equation 2}) and Eq.(\ref{equation 3}) with 1.96 TeV $p\bar{p}$
collision energy. Here we should clarify that the above defined
parton-level $\hat\theta$, which is the polar angle of outgoing
$e^-(e^+)$ with respect to the incoming parton from proton beam,
would be boosted to an experiment observable angle $\theta$
between $e^-(e^+)$ and proton beam. In this way, we get an event
generator of LFV $e\mu$ production at the Tevatron based on the
package Pythia. Now the question is, apart from the theoretical
calculation, how we can find some appropriate observables in which
the overlap of sneutrino and squark contribution parts is
minimized, so that one can suppress squark 'pollution' to a
trivial level and abstract sneutrino information clearly from
experiment.

\vskip 5mm
\section{Numerical results and discussion}

We take the R-parity violating parameters $\lambda$ and
$\lambda^{'}$ under the experimental constraints presented in
Ref.\cite{lambda1}. All the scale factors in the constraints which
are SUSY particle mass dependent are simply ignored, so that the
R-violating Yukawa couplings could be naturally kept at values
less than $\cal{O}$($10^{-1}$). Thereby, there is only mass scale
parameter $m_{\tilde{\nu}}$ in sneutrino section as free parameter
which should be determined by experiment. The degenerate sneutrino
mass parameter has to be constrained by the latest concrete LEP
collider data. For example, OPAL Experiment has set an upper limit
on $\sigma(e^+e^-\rightarrow e \mu)$ as $22fb$ with
$200\le\sqrt{s}\le 209GeV$ at 95\% CL \cite{LEPOPAL}. To get a
consistent value with this observation, one can arrive
at\cite{eeemJHEP}
\begin{equation}
\label{msnlimt} m_{\tilde\nu} \ge 250\mbox{GeV}.
\end{equation}
A unique sneutrino decay width $\Gamma$ is set to be $10GeV$,
which results in a very short $\tilde \nu$ lifetime. The
degenerate scalar quark mass is taken as $m_{\tilde q}=100GeV$
except otherwise specified, and $2 \times 2$ $R^{\tilde u}$ and
$R^{\tilde d}$ mixing matrices are set to be unit for
simplification.

\par
According to the decouple feature of sneutrino and squark section
at parton level, we plot two independent contributions to the
$e\mu$ signal production cross-section on $p\bar p$ collider at
$\sqrt s=1.96TeV$, as the functions of corresponding mass
parameter $m_{\tilde \nu}$ and $m_{\tilde q}$ respectively in
Figure 2. Due to the $\hat s$-channel resonance enhancement, the
sneutrino contribution is dominant over the squark one in most
region of parameter space. However, on account of the less density
of valence quark p.d.f in higher energy region, the resonance
enhancement would decrease with the increment of sneutrino mass,
and the cross section drops to a few femto barn in the vicinity of
$\sqrt{\hat s}\sim m_{\tilde{\nu}}=500$GeV. The calculation shows
as well that the cross-section at hadron level contributed by
scalar quark exchange $\hat t$+$\hat u$-channel is 5$fb$ when
$m_{\tilde{q}}=100$GeV, and becomes negligible when $m_{\tilde{q}}
\gg 100$GeV.

\par
A set of MC distributions of $e\mu$ signal observables at the
Tevatron, i.e. polar angle $cos\theta$, pseudo-rapidity
$\eta=-ln(\tan\frac{\theta}{2})$ and transverse momentum $p_T$ of
outgoing electron, and invariant mass of $e\mu$ system are plotted
in Figure 3, with $m_{\tilde{\nu}}=500$GeV and
$m_{\tilde{q}}=100$GeV where the pollution from scalar quark
section is maximized. The sneutrino contribution is characterized
by the di-lepton invariant mass peak around $M_{e\mu} = \sqrt{\hat
s} \sim m_{\tilde{\nu}}$, and the lepton transverse energy
tendency as $E_T \sim m_{\tilde{\nu}}/2$, while the two leptons
from squark section are much softer. Therefore, with a proper cut
on $e\mu$ invariant mass, we ought to get rid of squark pollution
and derive a purified sneutrino section. It should be clarified
here we adopt the Eq. (\ref{equation 2}) and Eq.(\ref{equation 3})
to generate the above two individual 'virtual' MC event samples
contributed by squark and sneutrino respectively, but would
generate the 'true' samples by using the differential
cross-section contributed by both squark and sneutrino sections in
following event selection strategy studies.

\par
For the background of the signal, we consider the following
processes which have significant $e\mu$ candidates in their final
states
$$
p \bar p \rightarrow W^+W^-, \tau^+\tau^-,b\bar b, t\bar t \rightarrow e\mu+X
$$
where $X$ refers to decay products of $W$,$\tau$,$b$ and $t$ other
than $e\mu$. As discussed in Ref\cite{eeemJHEP}, $b\bar b$ process
can be removed by calorimeter-based energy isolation cut on both
$e\mu$ candidates, for these leptons from b-quark semi-leptonic
decay are very soft and always associated with charmed meson as
$b\rightarrow qW^{\ast} \rightarrow jl\nu$. However, for $t\bar t$
events where top quark decay to $b W$ exclusively, there are possible
$e\mu$ final states from $WW$ subsequential decay, together with b-quarks
fragmentized into two jets which can not be tagged easily.
Different from $e^+e^-$ collider case, these $t\bar t$ production
events with $e\mu$ in final states accompanied by two b-quarks jets
might not be distinguished from those signal events, in which besides
$e\mu$ there are two jets produced by remnant parton collision or
multiple interactions in high luminosity hadron collider
environment. Therefore, there are three kinds of processes which
should be taken into account as physical background, and relevant
cross-sections at the Tevatron Run-II are
\begin{eqnarray}
\sigma_{WW} &=& \sigma[p\bar p\rightarrow W^+W^-] \times 2\cdot
         Br[W\rightarrow e\nu_e] \cdot Br[W\rightarrow \mu\nu_{\mu}]
         \nb \\
 &=& 8.45~pb \times 2\cdot 11\% \cdot 11\% = 204.5~fb\\
\nb\\
\sigma_{tt} &=& \sigma[p\bar p\rightarrow t\bar t] \times 2\cdot
         Br[W\rightarrow e\nu_e] \cdot Br[W\rightarrow \mu\nu_{\mu}] \nb\\
 &=& 8.32 pb \times 2\cdot 11\% \cdot 11\% = 201.3~fb\\
\nb\\
\sigma_{\tau\tau} &=& \sigma[p\bar p\rightarrow \tau^+ \tau^-]
\times 2\cdot
         Br[\tau \rightarrow e\nu_e\nu_{\tau}] \cdot
         Br[\tau \rightarrow \mu\nu_{\mu}\nu_{\tau}] \nb \\
 &=& 229.67~pb \times 2\cdot 17\% \cdot 17\% = 13.27~pb
\end{eqnarray}
Some MC distributions of $t\bar t$ and $\tau\tau$ background are
depicted in Figure 4 and Figure 5 respectively. Distributions of
$WW$ are very similar to those of $t\bar t$ and thus are not
plotted here. These distributions show that in $t\bar t$ and $WW$
events there are quite large missing transverse energy and
non-collinearity of $e\mu$ in azimuthal plane, while $e\mu$ final
states in $\tau\tau$ events are very soft and derive small
invariant mass. These features of background would be useful to
develop an efficient event selection strategy to suppress
background and improve the significance of signal.
\par
Taking the advantage of the collinearity and the resonance effect
of energetic $e\mu$ signal contributed by sneutrino section, we
suggest following off-line event selection strategy at
Run-II Tevatron to reduce SM physical background as well as squark
pollution
\begin{enumerate}
\item {\bf CUT1} Geometry Acceptance: Heavily boosted at the
Tevatron, the pseudo-rapidity distribution of signals does not
show distinct difference from those of background, namely all
$e\mu$ candidates would tend to forward-backward distributed.
Accordingly, we merely adopt the real detector acceptance on both
signal and background to avoid signal loss.
For demonstration, we take the real geometry of upgraded D\O\
detector\cite{D0detector} in simulation, where the pseudo-rapidity
coverage of Muon+Tracker system is as
\begin{eqnarray}
|\eta|\le 1.8~ \&~ NOT[|\eta|\le 1.25~ \& ~ 4.25 \le\phi\le 5.15]
\end{eqnarray}
and the coverage of the Central Calorimeter which provides better
energy resolution than the two Endcap ones is as
\begin{eqnarray}
|\eta| \le 1.2
\end{eqnarray}

\item {\bf CUT2} Transverse Momentum constraint: In order to
reject tremendous soft QCD jets which might produce counterfeit
'electron' and 'muon' simultaneously in high luminosity
hadron collision environment, we use $p_T$ cut on both electron
and muon candidates as:
\begin{eqnarray}0
p_T \ge 30~GeV
\end{eqnarray}

\item {\bf CUT3} Invariant Mass constraint: In order to remove the
scalar quark pollution from total signals, we make an invariant
mass cut as
\begin{equation}
M_{e\mu} \ge 140~GeV
\end{equation}
which is just greater than half of 250GeV, the current lower bound
on sneutrino mass given in Eq.(\ref{msnlimt}). We deliberately
choose such a loose cut to leave room for resolution and
uncertainty in real track transverse momentum and EM object energy
scale measurement. In this way, $\tau\tau$ events will be
suppressed heavily. Additionally, this invariant mass constraint
would also be efficient on some instrumental background, e.g.
single $Z^0$ production which decays via
$Z^0\rightarrow\mu^-\mu^+$ where one of its decay produced muons
occasionally deposits nearly all its energy in Calorimeter and
thus provides a fake electron candidate. However, a good-sized
part of $WW$ and $t\bar t$ events could survive after this cut yet.\\

\item {\bf CUT4} Collinearity constraint: Assuming high spatial
resolution on x-y plane vertical to the beam, we set a cut on
$e\mu$ $\phi$-difference to select back-to-back events as:
\begin{equation}
165^{o} \le \Delta \phi_{e\mu} \le 180^{o},
\end{equation}
where we allow a $15^{o}$ redundancy, considering that
the remnant parton collisions recoiled in signal processes might
carry some significant amount of transverse momenta in total.\\

\item {\bf CUT5} Missing Transverse Energy constraint: In LFV
signals there is not any missing $\rlap /E_T$. However, in
consideration of detector resolution we cut on a non-zero value as
\begin{equation}
\rlap /E_T \le 30~GeV.
\end{equation}

\end{enumerate}

\vpar Four sets of MC samples for $W^+W^-$,$\tau^+\tau^-$,$t\bar
t$ background and LFV signal productions with
$m_{\tilde{\nu}}=500$ and $m_{\tilde{q}}=100$GeV are generated by
Pythia at the Tevatron Run-II $p\bar p$ collider with
$\sqrt{s}=1.96~TeV$, scaled with different assumed luminosity.
Each sample is simulated with the five-step event selection, and
the numbers of events passing individual CUT are listed in Table
\ref{TableSB},respectively.
\begin{table}[htb]
\centering
\begin{tabular}{|l|c|c|c|c|}
\hline
                 &SM  background        & SM background              & SM background             & e-$\mu$ signal\\
                 & $W^+W^-\rightarrow e\mu$ & $\tau^+\tau^-\rightarrow
                 e\mu$ & $t\bar t\rightarrow e\mu$
                 &($m_{\tilde{\nu}}=500$GeV,\\
                 &  & & & ~~$m_{\tilde{q}}=100$GeV)\\
\hline
\hline
no cut $N_0$     & 9981 & 46066 & 9996 & 99952 \\
\hline
{\bf CUT1} $N_1$ & 5163 & 13575 & 6272 &36333 \\
\hline
{\bf CUT2} $N_2$ & 2312  & 140  & 3357 &31702 \\
\hline
{\bf CUT3} $N_3$ & 566  & 16    & 960 &26860 \\
\hline
{\bf CUT4} $N_4$ & 262  & 16    & 317 &26860  \\
\hline
{\bf CUT5} $N_5$ & 198  & 6     & 62 &26860 \\
\hline
efficiency $\epsilon$ & 2.0\% & 0.013\% & 0.6\% & 26.9\% \\
\hline\hline
$\sigma$ before CUT  & 204.5 $fb$ & 13.27 $pb$ & 201.3 $fb$ & 9.16 $fb$ \\
\hline
$\sigma$ after CUT   & 4.09 $fb$ & 1.73 $fb$ & 1.21 $fb$  & 2.46 $fb$ \\
\hline
\end{tabular}
\caption{\footnotesize Event selection efficiency on background and
signal. The first 6 rows are numbers of events before and after
individual CUT. The values of event selection efficiency on different
samples are given in the 7th row.}
\label{TableSB}
\end{table}
Despite different assumed luminosity in generating the MC samples,
we define a selection efficiency variable on both background and
signal as
\begin{eqnarray}
\epsilon=\frac{N_5}{N_0}
\end{eqnarray}
It's demonstrated that with the five-step strategy we are able to
reduce the background by 2-3 orders, i.e. $\epsilon_{WW}\sim
2.0\%$, $\epsilon_{t\bar t}\sim 0.6\%$ and
$\epsilon_{\tau\tau}\sim 0.013\%$, while keeping the selection
efficiency on the R-violating $e\mu$ signal as high as $27\%$.

\vpar
With certain integrated luminosity L, the number of background
events B and that of $e\mu$ signal S dominated by the contribution
of sneutrinos with $m_{\tilde \nu}$=500$GeV$ after selection are
given by
\begin{eqnarray}
S &=& (\sigma_{e\mu} \cdot \epsilon_{e\mu}) \cdot L
         = \sigma_{e\mu}^{CUT} \cdot L \nonumber\\
     &=&  2.46~fb \cdot L \\
\nb\\
B &=& (\sigma_{WW} \cdot \epsilon_{WW} +
            \sigma_{\tau\tau} \cdot \epsilon_{\tau\tau}+\sigma_{tt} \cdot \epsilon_{tt} ) \cdot L
        = \sigma_{SM}^{CUT} \cdot L
        \nonumber\\
        &=& 7.03~fb \cdot L
\end{eqnarray}
where $\sigma_{SM}^{CUT}$ is the sum of $W^+W^-$, $\tau^+\tau^-$
and $t\bar t$ contribution after CUTs, and $\sigma_{e\mu}^{CUT}$
is the cross-section of R-violating LFV $e\mu$ signal. The
significance of signal over background is defined as
\begin{equation}
\mbox{SB} = \frac{S}{\sqrt{B}}
       = \frac{\sigma_{e\mu}^{CUT}}{\sqrt{\sigma_{SM}^{CUT}}}
       \cdot \sqrt{L}
\end{equation}
In the following discussion, we optimistically assume some
10$fb^{-1}$ integrated luminosity would be accumulated at both the
D\O\ and CDF Experiment eventually, which is not an unreachable
goal for the Tevatron Run-II luminosity upgrading is in progress
remarkably. Accordingly, the significance of signal from 500 $GeV$
sneutrino contribution is about 3.

\vpar The significance $SB$ as functions of sneutrino mass
$m_{\tilde \nu}$ at the Tevatron is plotted in Figure 6, with
integrated luminosity being 5$fb^{-1}$ and 10$fb^{-1}$
respectively, where L=5$fb^{-1}$ is taken as conservative estimate
for reference. One can see clearly that LFV signals contributed by
sneutrinos lighter than 430$GeV$ would derive significance of
signal over the SM background greater than 5 for L=10$fb^{-1}$,
and thus would be detected explicitly. The signal cross-section
dominated by the contribution of 430$GeV$ sneutrinos would hold as
large as 4.2$fb$ after selection, and there will be some 40 signal
events on record together with 70 background events in total
$10fb^{-1}$ data. The transverse energy distributions of selected
electron candidates are plotted in Figure 7. Despite trivial
remnant of scalar quark pollution, the outstanding peak in higher
region of selected electron $E_T$ distribution denotes the half
value of sneutrino mass scale, and reveals the R-violating single
sneutrino production distinctively.

\par
Figure 6 and 7 show that with an optimized
10$fb^{-1}$ luminosity recorded in total, LFV signals of
degenerated sneutrinos with masses smaller than 400$GeV$ would be
detected at the Tevatron at more than 5 significance, and an
apparent precipitate structure shown in transverse energy
distributions of selected electron candidates would reveal the
whereabout of sneutrino mass scale. The ratio of signal to
background can be as large as 0.6, which is acceptable for a
stable data analysis. Consequently, if the sneutrino mass scale
really lies in the region of $m_{\tilde{\nu}}\le 400~GeV$, the
D\O\ Experiment at the Run-II Tevatron will discover these
supersymmetric scalar partners of Dirac neutrinos via LFV $e\mu$
signal processes; while if no experimental evidence is observed,
one can exclude sneutrino up to $m_{\tilde \nu}\ge 550~GeV$ at
95\% CL where significance is smaller than 2. We adopt the
criterion that the R-violating LFV signal is discovered at $SB>5$
and excluded when $SB<2$. Accordingly, Figure 8
shows the discovery and exclusion regions of sneutrino as
functions of integrated luminosity at the Tevatron. The solid
and dotted curves correspond to the significance of the
signal over background(SB) being $5$ and $2$, and thus the
sneutrino mass parameter regions above or under the hatched
area can be excluded or discovered, respectively.

\vskip 5mm
\section{Summary}

We have studied the lepton flavor violation processes $p\bar p \to
e\mu+X$ at the Tevatron Run-II with $\sqrt{s}= 1.96~TeV$ colliding
energy, in the framework of the R-parity violating MSSM. There are
two types of R-violating interactions involved in the LFV signal
processes, namely the couplings of sneutrino-leptons(quarks) vertices and
those of squark-quark-lepton ones. Fortunately, due to the R-violating S-P
Yukawa couplings of sfermion to Dirac leptons and quarks, the contribution of
sneutrino section is decoupled with that of squark part.
Accordingly, one is able to cut on an appropriate observable such
as invariant mass $M_{e\mu}$ to remove most of scalar quark
'pollution', and derive sneutrino mass parameter from purified
data sample which is dominated by sneutrino contribution.

\vpar Making use of the sneutrino resonance effect and $e\mu$
collinearity in transverse plane vertical to beam pipe, we develop
a set of event selection strategies. Under these strategies, the
SM background can be under control, so that it is possible to
probe the R-parity violating interactions and to abstract the
sneutrino information from experimental observation. We conclude
that with an assumption of 10$fb^{-1}$ integrated luminosity, the
Experiments at the Tevatron Run-II machine would have the
potential to discover sneutrino in the region of
$m_{\tilde{\nu}}\le 400 GeV$ by detecting LFV $e\mu$ signals, or
extend the mass scale constraint up to $m_{\tilde{\nu}}\ge
550~GeV$ at 95\% CL within the MSSM framework with R-parity
violation.

\vskip 5mm \noindent{\large {\bf Acknowledgement:}} This work was
supported in part by the National Natural Science Foundation of
China and special fund sponsored by China Academy of Science.

\vskip 10mm
\begin{flushleft} {\bf Figure Captions} \end{flushleft}
\vskip 6mm

\par{\bf Fig.1} The Feynman diagrams of subprocess $q\bar{q}\rightarrow e^\mp\mu^\pm$.

\par{\bf Fig.2} The $e\mu$ signal production cross section parts contributed by sneutrino
                and squark sectors respectively, as the functions of $m_{\tilde \nu}$ and
                $m_{\tilde q}$ at the Tevatron Run-II.

\par{\bf Fig.3} MC distributions of $e\mu$ signal at the Tevatron Run-II. The hatched histograms
                are from squark section, and the unhatched histograms are given by squark+sneutrino
                section contribution.

\par{\bf Fig.4} MC distributions of the $tt$ background.

\par{\bf Fig.5} MC distributions of the $\tau\tau$ events background.

\par{\bf Fig.6} The significance SB as a function of $M_{\tilde \nu}$ with $M_{\tilde q}=100GeV$
                at the Tevatron with different integrated luminosity estimates.

\par{\bf Fig.7} distributions of electron candidates after selection. The numbers are
                normalized with integrated luminosity $10fb^{-1}$. The hatched histogram
                is for SM background, and the unhatched one is background folded with 430GeV
                sneutrino signal.

\par{\bf Fig.8} Discovery(SB=5, solid line) and exclusion(SB=2, dot line) limits
                on sneutrino mass parameter as a function of the Tevatron integrated
                luminosity. The region between these two lines are those which can not
                be excluded or discovered explicitly.

%%%%%%%%%%%%%%%%%%%%%%%%%%%%%%

\vskip 10mm
\begin{thebibliography}{s25}
\bibitem{neutrinoexp}
    Y. Fukuda et al., Super-Kamiokande collaboration, Phys. Lett. {\bf B433} (1998)9;
    Phys. Lett. {\bf B436} (1998)33;  Phys. Rev. Lett. {\bf 81} (1998)1562; 
    M. Apollonio et al., Chooz collaboration, Phys. Lett. {\bf B420} (1998)297.

\bibitem{MSSM}
    H. P. Nilles, Phys. Rep. {\bf 1},(1984)110; 
    H.E. Haber and G. Kane, Phys. Rep. {\bf 117},(1985) 75.

\bibitem{R} 
    P. Fayet, Phys. Lett. {\bf B 69} (1977) 489; 
    G. R. Farrar and P. Fayet, Phys. Lett. {\bf B 76} (1978) 575.

\bibitem{potential1} 
    S. Weinberg, Phys. Rev. {\bf D26} (1982) 287;
    N. Sakai, T. Yanagida, Nucl. Phys. {\bf B197} (1982) 533.

\bibitem{Rbilinear} 
    J. C. Romao, M. A. Diaz, M. Hirsch, W. Porod and J. W. Valle, Phys. Rev. {\bf D 61} (2000)071703; 
    Phys. Rev. {\bf D62} (2000)113008.

\bibitem{LEPOPAL}
    G.Abbiendi {\it et al}., Phys Lett. {\bf B519}(2001)23-32.

\bibitem{eeemJHEP}
    Y.B. Sun, L. Han, W.G. Ma, T. Farshid, R.Y. Zhang and Y.J. Zhou, JHEP 0409 (2004) 043.

\bibitem{sneu} 
    R. Barbieri et al., hep-ph/9810232; B. Allanach et al., hep-ph/9906224;
    F. Deliot et al, Phys. Lett.{\bf B475} (2000)184;
    G. Moreau et al. Nucl. Phys. {\bf B604} (2001)3;
    S. Bar-Shalom et al. Phys. Rev. {\bf D64} (2001) 095008.

\bibitem{pythia}
    T. Sj$\ddot{o}$strand, P. Ed$\acute{e}$n, C. Friberg, L. L$\ddot{o}$nnblad, G. Miu, S. Mrenna and E. Norrbin,
    Computer Phys. Commun. 135 (2001) 238 (LU TP 00-30, [hep-ph/0010017])

\bibitem{lambda1}
    R. Barbieri {\it et al}, [hep-ph/9810232];
    B. Allanach {\it et al}, [hep-ph/9906224].

\bibitem{D0detector} 
    The D\O\ detector is described in detail elsewhere as:
    T.LeCompte and H.T.Diehl, "The CDF and D\O\ Upgrades for Run II'',  Ann. Rev. Nucl. Part. Sci. {\bf 50}, 71 (2000);
    V. Abazov, {\sl et al.}, in preparation for submission  to Nucl. Instrum. Methods Phys. Res. A.

\end{thebibliography}
\end{document}